\documentstyle[11pt,newpasp,twoside,epsf]{article}
\def\edcomment#1{\iffalse\marginpar{\raggedright\sl#1\/}\else\relax\fi}
\reversemarginpar

\begin{document}
\title{Neutron Stars in Supernova Remnants and Beyond}
 \author{Vasilii V. Gvaramadze}
\affil{Sternberg Astronomical Institute, Moscow State University,
Universitetskij Pr.~13, Moscow, 119992, Russia}

\begin{abstract}
We discuss a concept of off-centred cavity supernova explosion as
applied to neutron star/supernova remnant associations and show
how this concept could be used to preclude the anti-humane
decapitating the Duck (G\,5.4$-$1.2 $+$ G\,5.27$-$0.9) and
dismembering the Swan (Cygnus Loop), as well as to search for a
stellar remnant associated with the supernova remnant RCW\,86.
\end{abstract}

\section{Introduction}

Massive ($\geq 8-10 \, M_{\odot}$) stars are the progenitors of
most of supernovae (SNe). The explosion of a massive star results
in the origin of an extended (tens of parsecs) diffuse supernova
remnant (SNR) and a compact stellar remnant. In most cases the
stellar remnant is a neutron star (NS). Therefore most of SNRs
should be associated with NSs. A NS could be located within the
confine of the associated SNR or beyond it, depending on the kick
velocity received by the star, the age of the system and some
other factors (see below). However, only a small fraction of known
SNRs was found to be associated with NSs. Moreover, it is believed
that some of proposed NS/SNR associations are the result of
geometrical projection. Thus, the obvious lack of associations
should be explained or filled up. The latter seems to be more
attractive and fruitful in view of the recent splash of
discoveries of NSs in SNRs.

The reliability of NS/SNR associations is usually assessed with
help of several criteria (see e.g. Kaspi 1996). Some of them are
trivial, i.e. should be fulfilled for any association. Other
criteria are based on the use of the standard Sedov-Taylor model
of SNRs and therefore are over-simplified. The application of
these criteria can lead to rejection of genuine associations
(Gvaramadze 2000, 2002a,b; Bock \& Gvaramadze 2002). The point is
that the massive stars strongly modify their environs by virtue of
their winds and ionizing emission, and it is the subsequent
interaction of SN blast waves with their processed ambient medium
that results in the observed SNRs (e.g. Shull et al. 1985; Ciotti
\& D'Ercole 1989; Chevalier \& Liang 1989). It is clear that the
presence of circumstellar and interstellar structures could
strongly affect the standard sequence and duration of evolutionary
stages of the SN blast wave (e.g. Woltjer 1972). For example, the
Sedov-Taylor stage could be absent at all if the SN exploded
within the wind-blown cavity surrounded by a massive shell (e.g.
Franco et al. 1991). Another important point is the proper motion
of massive stars, which causes them to explode far from the
geometrical centres of their cavities and makes the cavities and
other circumstellar and interstellar structures
non-spherically-symmetrical. The natural consequence of an
off-centred cavity SN explosion is that the SN blast centre does
not coincide with the geometrical centre of the future SNR. Taking
into account these considerations allows us to enlarge the circle
of possible NS/SNR associations and to explain the morphological
peculiarities of SNRs (Gvaramadze 2000, 2002a,b; Bock \&
Gvaramadze 2002; Gvaramadze \& Vikhlinin 2003). On the other hand,
the better understanding of origin of peculiar SNRs helps to infer
the ``true" SN explosion sites in these remnants, and therefore to
search for new NSs associated with them (Gvaramadze 2002b;
Gvaramadze \& Vikhlinin 2002, 2003; see also Sect.\,4).

To illustrate the significance of the concept of off-centred
cavity SN explosion we show how this concept could be used to
preclude the anti-humane decapitating the Duck (G\,5.4$-$1.2 $+$
G\,5.27$-$0.9) and dismembering the Swan (Cygnus Loop,
G\,74.0$-$8.5) recently attempted, respectively, by Thorsett,
Brisken, \& Goss (2002) and Uyaniker et al. (2002), and to search
for a NS associated with the SNR RCW\,86 (MSH\,14-6{\it3},
G\,315.4$-$2.30).

\section{The Duck (G\,5.4$-$1.2 $+$ G\,5.27$-$0.9)}

The association between the pulsar B\,1757$-$24 and the SNR
G\,5.4$-$1.2 was for a long time one of a few most reliable NS/SNR
associations. However, recently this association was questioned by
Thorsett et al. (2002), who suggested that PSR B\,1757$-$24 and
the compact nebula G\,5.27$-$0.9 behind it (the ``head" of the
``Duck") are unrelated to the SNR G\,5.4$-$1.2 (the ``body" of the
``Duck"). To justify the decapitating the Duck, Thorsett et al.
used a set of standard criteria for evaluating the reliability of
NS/SNR associations (e.g. Kaspi 1996), which include the agreement
of distance and age estimates for pulsar and SNR, the consistence
of the implied pulsar transverse velocity (i.e. the velocity
inferred by the displacement of the pulsar from the geometrical
centre of the SNR) with the measured (e.g. proper motion)
velocity, and the correct orientation of the vector of pulsar
transverse motion. We agree with Thorsett et al. (2002) that the
estimates of the distance to the pulsar and the SNR are not
inconsistent and therefore will concentrate on the rest three
criteria.

Two main arguments against the association put forward by Thorsett
et al. (2002) are based on their interferometric proper motion
measurements of PSR B\,1757$-$24 [an upper limit on the westward
motion of the pulsar, $v_{\rm w}$ $\leq 145 \, d_{4.5}$ $\, {\rm
km} \,$ ${\rm s}^{-1}$ ($d_{4.5}$ is the distance to the SNR in
units of 4.5 kpc) was found to be an order of magnitude less than
the implied transverse velocity] and the ``incorrect" orientation
of the pulsar proper motion (a cometary-shaped nebula behind the
pulsar does not point back to the geometrical centre of
G\,5.4$-$1.2). Both ``inconsistencies", however, could be removed
if the SNR G\,5.4$-$1.2 is the result of off-centred cavity SN
explosion. We suggest that: a) PSR B\,1757$-$24 was born near the
northwest edge of a wind-blown cavity; b) the cavity was
surrounded by a massive ($\geq 50 \, M_{\rm ej}$, where $M_{\rm
ej} \simeq 3-4 \, M_{\odot}$ is the mass of the SN ejecta)
wind-driven shell (see also Gvaramadze 2000, 2002b). The first
suggestion implies that: {\it i}) the ``true" transverse velocity
of the pulsar is much smaller than the implied one; {\it ii}) the
tail behind the pulsar has a correct orientation. The second
suggestion implies that: {\it i}) the SN blast wave was
drastically decelerated by the interaction with the wind-driven
shell, so that the pulsar (moving in the westward direction at the
velocity $\leq v_{\rm w}$) was able to overrun the resulting SNR;
{\it ii}) the current radius of the SNR is about the same as the
radius of the wind-driven shell.

The proper motion measurements by Thorsett et al. (2002) provide
an estimate of the kinematic age of the system: $t_{\rm kin} \sim
l/v_{\rm w}$, where $l$ is the distance travelled by the pulsar
from its birthplace. For $l\simeq 8 \, d_{4.5}$ pc (Gvaramadze, in
preparation), one has $t_{\rm kin} \geq 3\tau$, where $\tau =1.55
\times 10^4 \, {\rm yr}$ is the characteristic age of the pulsar.
The ``true" pulsar age, however, could be equal to $t_{\rm kin}$
if the spin-down rate of the pulsar is mainly due to the
interaction between the pulsar's magnetosphere and the dense
ambient medium (see Gvaramadze 2001 and references therein), or if
the pulsar braking index is $\leq 1.7$ (cf. Gaensler \& Frail
2000).

\section{The Swan (Cygnus Loop, G\,74.0$-$8.5)}

Recent polarized intensity image of the Cygnus Loop obtained by
Uyaniker et al. (2002) revealed a prominent shell-like structure
encompassing the ``break-out" region in the south of this SNR.
Uyaniker et al. advocated the widely accepted point of view on the
origin of SNRs consisting of two overlapping shells and suggested
that the Cygnus Loop is actually two individual SNRs interacting
with each other. An alternative possibility is that the SNRs of
this type are due to off-centred cavity SN explosions (Dubner et
al. 1994, Gvaramadze 2002b). We suggest that the SNR Cygnus Loop
is the result of SN explosion near the south edge  of a cavity
blown up by the SN progenitor during the main-sequence phase (cf.
Gvaramadze \& Vikhlinin 2003; see also Sect.\,4). This implies
that the conventional shell of the Cygnus Loop corresponds to the
former cavity re-energized by the SN blast wave, while the south
shell is created by the interaction of the SN blast wave with the
unperturbed interstellar medium. Accordingly, we expect that only
one stellar remnant is associated with both shells.

Uyaniker et al. (2002) discussed eight reasons for considering the
Cygnus Loop as two colliding SNRs. We note however that these
reasons could also be considered as indications of the off-centred
cavity SN explosion. For example, we believe that the presence of
a NS candidate (Miyata et al. 2001) near the centre of the south
shell is a strong argument in support of our suggestion. Note also
that the absence of centrally peaked X-ray emission in the south
shell implies that the SN progenitor exploded after the red
supergiant phase of its evolution (cf. Gvaramadze 2002b), i.e. the
initial mass of the progenitor was $< 15\, M_{\odot}$.

\section{RCW\,86 (MSH\,14-6{\it 3}, G\,315.4$-$2.30)}

RCW\,86 is a bright shell-like SNR with a peculiar protrusion in
the southwest encompassing a prominent hemispherical optical
nebula. We believe that RCW\,86 is the result of a cavity SN
explosion of a moving massive star, which after the main-sequence
phase has evolved through the red supergiant phase, and then
experienced a short ``blue loop" (Gvaramadze \& Vikhlinin 2003;
cf. Gvaramadze 2002b). During the main-sequence phase the stellar
wind blows up a large-scale cavity, while the motion of the star
causes it to cross the cavity and start to interact directly with
the interstellar medium. We suggest that the SN exploded inside a
``hollow" bow shock-like circumstellar structure (created by the
post-main-sequence winds) adjacent to the main-sequence cavity.
This suggestion implies that the optical arc in the southwest of
RCW\,86 is the remainder of the pre-existing circumstellar
structure and that the stellar remnant associated with the SNR
should be in the centre of the arc.

Motivated by these considerations we searched for a stellar
remnant in the southwest protrusion of RCW\,86 using the {\it
ROSAT} and {\it Chandra} archival data. The unprecedented high
angular resolution of the {\it Chandra X-ray Observatory} allows
us to detect a NS candidate just in the ``proper" place (see
Gvaramadze \& Vikhlinin 2002, 2003).

\end{document}